\documentclass[aps,twocolumn,prb,groupedaddress,slide,nofootinbib]{revtex4-1}

\usepackage{amsmath}
\usepackage{braket}
\usepackage{xspace}
\usepackage{graphicx}

\begin{document}

\draft

\title{Analyzing scanning tunneling spectroscopy for Fe-based superconductor Ba$_{1-x}$K$_x$Fe$_2$As$_2$ and extracting $s$-wave density of states}
\author{Jongbae Hong}
\affiliation{Institute for Basic Science, Incheon National University,
 Yeonsu-gu, Incheon 22012, Korea}
\date{\today}

\begin{abstract}

We extract the density of states (DOS) from the scanning tunneling spectroscopy data for
Ba$_{1-x}$K$_x$Fe$_2$As$_2$ superconductor.
The obtained sample DOS is composed of two ordinary $s$-wave types from the band at $\Gamma$ point
and a linear-like DOS within the $s$-wave gap from the band at M point in the Brillouin zone, and
is consistent with the corresponding data from angle-resolved photoemission spectroscopy.
We clarify that the major peak of the tunneling conductance is not related to the DOS
but is rather the effect of nonequilibrium coherent tunneling including all coherent spins in the tip
and sample.

\end{abstract}

\maketitle \narrowtext

\section{Introduction}
The high transition temperature observed in single-layer FeSe on SrTiO$_3$ (FeSe/STO)~\cite{Ge,Wang}
discovered recently has further heightened interest in Fe-based superconductors.
These superconductors have multiple Fermi surfaces around $\Gamma$ and M points
in the Brillouin zone,~\cite{Hoffman} even though the band at the $\Gamma$ point is largely suppressed
below the Fermi level in single-layer FeSe/STO.~\cite{Lee,Liu}
Such multiplicity of Fermi surfaces raises a big challenge not only to identify the sample
density of states (DOS) but also to explain the tunneling
conductance line shape as measured by scanning tunneling spectroscopy (STS).~\cite{Wray,Chi}

Fundamental information on exotic superconductors is usually obtained by
two spectroscopic tools, namely angle-resolved photoemission spectroscopy (ARPES) and STS.
But a clearly known fact is that the proposed superconducting gaps by ARPES and STS
differ on a meaningful scale;
this is a long-standing puzzle in studying both cuprate~\cite{Lawler,Pushp,Shen}
and Fe-based superconductors.~\cite{Wray,Chi,Umezawa}
Another aspect of the Fe-based superconductor is the non-observation of $s$-wave DOS
in the tunneling conductance,~\cite{Wray,Chi}
even though the material has been claimed to be an $s^\pm$-wave superconductor.~\cite{Bang}
These two particular aspects---the inconsistency between ARPES and STS and
the lack of $s$-wave DOS in the tunneling conductance---are important issues which must be clarified.

Since a photon does not experience Coulomb interaction in a collision with correlated material,
we expect that the ARPES setup does not encounter any theoretical difficulty in explaining its output.
However, when an electron enters a correlated sample, strong Coulomb repulsion plays a crucial
role in the subsequent tunneling dynamics.
The STS setup sketched in Fig.~1(a) therefore requires a sophisticated theoretical treatment
because the coherent tunneling here occurs in a many-body {\it entangled state} involving
all coherent spins in the tip and sample.
The purpose of this study is to solve the two problems mentioned above by explicitly
obtaining the sample DOS from existing STS data and analyzing the contributions from different Fermi surfaces.

Among the many Fe-based superconductors, we focus on Ba$_{1-x}$K$_x$Fe$_2$As$_2$~\cite{Wray}
and single-layer FeSe/STO,~\cite{Wang}
because the former has contributions from bands at both $\Gamma$ and M points while
the latter is known to have contributions mostly from a band at the M point.~\cite{Lee,Liu}
Therefore, a comparative analysis of the STS data of these two materials
will be helpful to understand the nature of Fe-based superconductors.

\begin{figure}
[b] \vspace*{3.4cm} \includegraphics{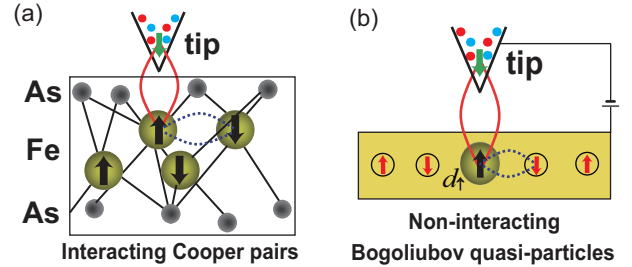} \vspace*{0.0cm}
\caption{(a) An STS setup for an Fe-based superconductor depicting an entangled state connected by
linearly combined singlets (red and dotted loops).
The red and blue circles in the tip denote coherent electrons with different spins,
and the arrows in the superconductor represent spins forming Cooper pairs.
(b) Description of the superconductor by means of non-interacting Bogoliubov quasiparticles ($T=0$ K).
The operator $d_{\uparrow}$ denotes up-spin annihilation at the MS.
}
\end{figure}

The prevailing interpretation of tunneling conductance, which is believed as the sample DOS, is
valid for materials without correlation effects.
However, when tunneling occurs in a correlated material with a bias
weaker than on-site Coulomb repulsion, an electron cannot simply hop to the mediating
site (MS) under the tip if an electron already occupies the MS.
In such a case, the only possible way for a coherent current to flow is to form a linearly combined
singlet connecting the tip and sample, as shown in Fig.~1(a), and perform a process of
singlet co-tunneling, changing singlet partner, and then co-tunneling again, repeatedly.
In this circumstance, the entering electron does not feel strong Coulomb repulsion at the MS.
Here, all coherent spins in the tip and all Cooper pairs in the sample are involved in
coherent tunneling by the linearly combined two singlets connecting the tip, MS,
and sample; we call this entangled state tunneling.
Many incoherent spins also exist in the STS system, and even though they do not join in
coherent current they influence current flow by causing double occupancy at the MS.
Accordingly, the effects of both coherent and incoherent spins should be properly treated
in related tunneling theory.

This study is designed as follows.
We introduce a tunneling conductance formula and the Green's function technique in Liouville space in order to
construct a simplified theoretical model appropriate to describe the STS system
under consideration in Section II.
In Section III, we determine the basis operators spanning the working Liouville space, and
present an explicit expression of nonequilibrium expectation for the entangled state tunneling
under consideration. Then, we prove the orthogonality of the basis operators.
In Section IV, we obtain a Liouville matrix and express the tunneling conductance as a function of sample DOS.
We present the results of fitting the experimental tunneling conductance of single-layer FeSe/STO~\cite{Wang}
and Ba$_{1-x}$K$_x$Fe$_2$As$_2$~\cite{Wray} with our extracted sample DOS in Section V, where
we explicitly show that the sample DOS is indeed composed of ordinary $s$-wave DOS coming from the band
at the $\Gamma$ point and a linear-type DOS from the band at the M point.
Our work concludes with discussion in Section VI.

\section{Green's function via Liouville approach}
We employ a tunneling conductance formula derived from the Meir--Wingreen current formula~\cite{Meir-PRL68}
\begin{equation}
    \frac{dI}{dV} = \left. \frac{e^2}{\hbar} \tilde{\Gamma}(\omega)
    \rho_d(\omega) \right|_{\hbar\omega=eV},
	\label{eq:dIdV}
\end{equation}
where $I$ and $V$ are current and bias voltage between tip and sample,
$\tilde{\Gamma}(\omega)=\Gamma^T\Gamma^S(\omega)/[\Gamma^T+\Gamma^S(\omega)]$
is the effective coupling of the tip, MS, and sample, and
$\Gamma^S(\omega)=2\pi(\tilde{V}^S)^2D(\omega)$, where $D(\omega)$ denotes the sample DOS.
Equation (\ref{eq:dIdV}) is obtained under the conditions $\Gamma^T(\omega_i\pm eV)\propto\Gamma^S(\omega_i)$
for each $\omega_i$ and bias independence of local DOS $\rho_d(\omega)$ that is different from
the sample DOS $D(\omega)$.
These conditions are satisfied in entangled state tunneling as mentioned above.
The key quantity of Eq. (\ref{eq:dIdV}) is the local DOS at the MS, which is given by
the Green's function at the MS: $\rho_d(\omega)=-(1/\pi){\rm Im}{\mathcal G}_{dd\sigma}^+(\omega)$.
This Green's function ${\mathcal G}_{dd\sigma}^+(\omega)$ should not be obtained at equilibrium but
rather at steady-state nonequilibrium.

We employ the resolvent Green's function in the operator space, called the Liouville space,
\begin{equation}
	{\mathcal G}^{+}_{pq}(\omega) = \langle{\hat e}_p|
	[(\omega+i0^+){\bf I}-{\bf L}]^{-1}|{\hat e}_q\rangle,
\label{eq:GRL}
\end{equation}
where ${\hat e}_q$ is the basis operator of the Liouville
space, ${\bf I}$ is the identity operator, and ${\bf L}$ is the
Liouville operator defined by ${\bf L}\cal O\equiv{\cal H}\cal O-\cal
O\cal H$ for operator $\cal O$.
The inner product in the Liouville space is defined by $\langle \cal A|\cal B\rangle\equiv\langle
\cal A\cal B^\dagger+\cal B^\dagger \cal A\rangle$, where $\cal B^\dagger$ is
the adjoint of $\cal B$ and the angular brackets denote the expectation value at steady-state nonequilibrium.
Equation (\ref{eq:GRL}) is a generic form of Green's function along with one using a Hamiltonian.
The reason we employ the Liouville approach instead of the well-known Hamiltonian approach is
the availability of basis operators;~\cite{ijmp2017,Hong-JPCM23a,Hong-JPCM23b}
the basis state vectors spanning the Hilbert space of a correlated system are not easily obtained.

The STS setup of Fig.~1(a) is described by the theoretical model
\begin{eqnarray}
	{\cal H} &=& \sum_{p,\sigma}\varepsilon_p^Tc_{p\sigma}^{T\dagger}c^T_{p\sigma}
+{\tilde V}^T\sum_{p,\sigma}(c_{p\sigma}^{T\dagger}d_{\sigma}+d_{\sigma}^\dagger c^T_{p\sigma})
+Un_{d\uparrow}n_{d\downarrow}\nonumber \\
&+&\sum_{k,\sigma}\varepsilon_k^Sc_{k\sigma}^{S\dagger}c^S_{k\sigma}+\sum_{kk'}M_{k,k'}
c_{k\uparrow}^{S\dagger}c_{-k\downarrow}^{S\dagger}c_{k'\uparrow}^Sc_{-k'\downarrow}^S,
\label{Hamil}
\end{eqnarray}
where the superscripts $T$ and $S$ denote the tip and superconductor sample, respectively.
We perform the Bogoliubov transformation for the fermion operators of superconductors using
$c_{k\uparrow}^{S\dagger}=u_k\alpha_{k\uparrow}^\dagger+v_k\alpha_{-k\downarrow}$ and
$c_{-k\downarrow}^{S\dagger}=u_k\alpha_{-k\downarrow}^\dagger-v_k\alpha_{k\uparrow}$.
Then, the second line of Eq. (\ref{Hamil}) changes to
$E_0+\sum_{k}E_k(\alpha_{k\uparrow}^\dagger\alpha_{k\uparrow}+\alpha_{-k\downarrow}^\dagger\alpha_{-k\downarrow})$
by neglecting quasiparticle interaction, which is valid at zero temperature,
where $E_0$ is the ground state energy of the superconductor and $E_k$ is the excitation energy
of the Bogoliubov quasiparticles.~\cite{note}
Then, the Hamiltonian of Eq. (\ref{Hamil}) is transformed into a two-reservoir quantum impurity system, as
\begin{eqnarray}
	{\cal H} &=& \sum_{p,\sigma}\varepsilon_p^Tc_{p\sigma}^{T\dagger}c^T_{p\sigma}
+{\tilde V}^T\sum_{p,\sigma}(c_{p\sigma}^{T\dagger}d_{\sigma}+d_{\sigma}^\dagger c^T_{p\sigma})
+Un_{d\uparrow}n_{d\downarrow}\nonumber \\
&+&{\tilde V}^S\sum_{k}(\alpha_{k\uparrow}^{\dagger}d_{\uparrow}+d_{\uparrow}^\dagger \alpha_{k\uparrow}
+\alpha_{-k\downarrow}^{\dagger}d_{\downarrow}+d_{\downarrow}^\dagger \alpha_{-k\downarrow})\nonumber \\
&+&\sum_{k}E_k(\alpha_{k\uparrow}^\dagger\alpha_{k\uparrow}+\alpha_{-k\downarrow}^\dagger\alpha_{-k\downarrow}),
\label{Hamil2}
\end{eqnarray}
which is depicted in Fig.~1(b).
This system has a flat DOS for the tip and a DOS given by $D(\omega)=\sum_k\delta(\omega-E_k)$
for the sample. Our main purpose is to find $D(\omega)$ from tunneling data.

\begin{figure}
[t] \vspace*{6.5cm} \includegraphics{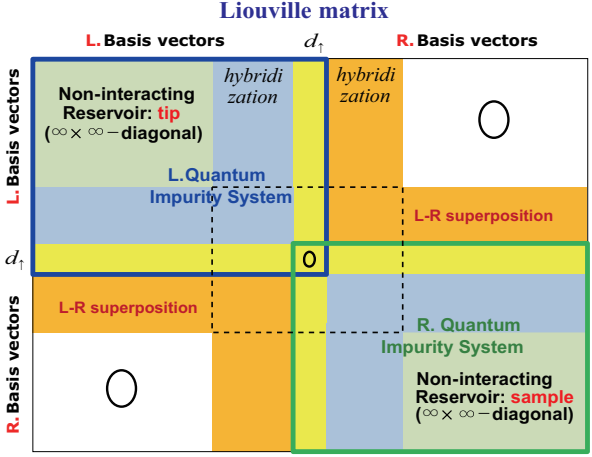} \vspace*{0.0cm}
\caption{
    Structure of the Liouville matrix for the system given in Fig.~1(b). The left and right sides represent the tip and
    superconductor sample, respectively. Each side is described by basis operators
    indicating the quantum states of the reservoir, hybridization, and the MS ($d_{\uparrow}$).
    The corner blocks are composed of tip--sample superposition and null blocks.
The Liouville matrix is reduced to the region indicated by the dashed line. }
\end{figure}

To calculate the Green's function of Eq. (\ref{eq:GRL}), the first step is to determine the complete set of basis operators.
One advantage of using the Liouville approach is that, as shown with different colors in Fig.~2,
the basis operators are clearly classified into three subgroups that play distinct roles:
the reservoir (light green), the MS (yellow), and hybridization (light blue).
Then, the Liouville matrix has an important section representing tip--sample coherence
at both corners of the matrix, as shown by the orange regions in Fig.~2.

For up-spin dynamics $d_{\uparrow}(t)$, the MS is described by $d_{\uparrow}$ solely due to strong on-site Coulomb repulsion
at the MS, and the reservoir is described by $c^T_{k\uparrow}$ where $k=1, 2, \cdots, \infty$.
A nontrivial part of the basis operators is that describing hybridization, which also has infinite degrees of freedom.
Fortunately though, the property of unidirectional movement in entangled state tunneling
restricts multiple back-and-forth processes, and this greatly simplifies the degrees of freedom for hybridization.
Thus, the Liouville matrix has infinite-dimensional parts only in the sectors describing the non-interacting reservoirs,
as shown by the light green regions in Fig.~2.
Since these non-interacting reservoirs (tip and sample) are represented by diagonal blocks,
the Liouville matrix can be reduced to a finite-dimensional matrix (indicated by the dashed line
in Fig.~2) by way of a matrix reduction scheme.~\cite{ijmp2017,Loewdin,Mujica}
Such matrix reduction corresponds to tracing out the degrees of freedom of the non-interacting reservoirs.

\section{Orthonormal basis operators and Nonequilibrium expectation}
Specifically for $d_{\uparrow}(t)$ dynamics, the MS is described by $d_{\uparrow}$ only,
the tip (sample) is described by $c^T_{k\uparrow} (\alpha_{k\uparrow})$ and
$n_{d\downarrow}c^T_{k\uparrow} (n_{d\downarrow}\alpha_{k\uparrow})$ where $k=1, 2, \cdots, \infty$,
and the hybridization is described by $(j^{T,S\pm}_{d\downarrow})d_{\uparrow}$
where $j^{T+}_{d\downarrow}={\tilde V}^T\sum_k(c^{T\dagger}_{k\downarrow} d_{\downarrow}+d^\dagger_{\downarrow} c^T_{k\downarrow})$ and
$j^{S-}_{d\downarrow}=i{\tilde V}^S\sum_k(\alpha^{\dagger}_{-k\downarrow} d_{\downarrow}-d^\dagger_{\downarrow} \alpha_{-k\downarrow})$.
The working Liouville space for the entangled state tunneling in STS is spanned by
the following basis operators divided into three groups:
\begin{alignat*}{4}
	& \{c_{k\uparrow}^T, \,\, c_{k\uparrow}^T\delta n_{d\downarrow}/\langle(\delta n_{d\downarrow})^2\rangle^{1/2} \}, & &
		k=0, 1, \ldots, \infty; \\
	& \{d_{\uparrow}\delta j^{-T}_{d\downarrow}, \,\,
		d_{\uparrow}\delta j^{+T}_{d\downarrow}, \,\,
		d_{\uparrow}, \,\, d_{\uparrow}\delta j^{+S}_{d\downarrow}, \,\,
		d_{\uparrow}\delta j^{-S}_{d\downarrow}\}; & & \\
	& \{\alpha_{k\uparrow}\delta n_{d\downarrow}/\langle(\delta n_{d\downarrow})^2\rangle^{1/2}, \,\, \alpha_{k\uparrow}\}, &
		& k=0, 1, \ldots, \infty,
\end{alignat*}
where we introduced $\delta$ indicating $\delta{\cal O}={\cal O}-\langle {\cal O}\rangle$ to achieve orthogonality among the basis operators.
We omit normalization factors $\langle(\delta j^{\pm T,S}_{d\downarrow})^2\rangle^{1/2}$
in the denominators of the corresponding basis operators in the second line.
Orthogonality between an operator without $\delta$ and one with $\delta$ is obvious;
however, proving orthogonality among operators containing $\delta$ is nontrivial.
Orthogonality for the latter case is clarified after determining the expectation
scheme for entangled state tunneling at steady-state nonequilibrium below.

\begin{figure}
[t] \vspace*{5cm} \includegraphics{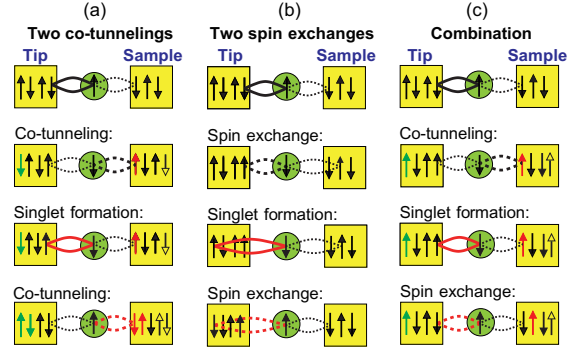} \vspace*{0.0cm}
\caption{ Coherent spin dynamics in an STS system.
The first row indicates the initial entangled state, and the third row indicates singlet partner change.
Red arrows indicate spin transferred from the tip to the sample.
Open and green arrows denote the spins leaving and entering the reservoir for steady current, respectively.
(a) Two-electron transfer via two singlet co-tunnelings.
(b) Kondo coupling dynamics on the left side.
(c) One-electron transfer via the combination of spin exchange and singlet co-tunneling.
}
\end{figure}

In STS of a correlated system in which both tip and sample are within the coherent region,
tunneling must be entangled state tunneling, as depicted in Fig.~1(b).
Under this circumstance, three different types of coherent processes as sketched in Fig.~3
are considered: (a) one comprising two singlet co-tunnelings,
(b) one comprising two spin exchanges describing Kondo coupling, and
(c) one comprising spin exchange and singlet co-tunneling.
Figure 3(a) transports two electrons (red arrows) after one cycle,
Fig.~3(b) represents Kondo coupling dynamics on the left side,
and Fig.~3(c) transports one electron (red arrow) after one cycle.
One cycle thus comprises four hybridization processes.

The steady-state nonequilibrium expectation for Fig.~3(a) is expressed by
\[\langle\Psi_0|{\tilde V}^S\alpha_{k\uparrow}^{\dagger}(c_{k\uparrow}^T d_\uparrow^\dagger \alpha_{-k\downarrow}^{\dagger}
d_\downarrow d_\downarrow^\dagger c_{k\downarrow}^T) d_\uparrow|\Psi_0\rangle/\Delta_0, \] while for Fig.~3(c) by
\[\langle\Psi_0|{\tilde V}^S\alpha_{k\uparrow}^{\dagger}(c_{k\uparrow}^T d_\uparrow^\dagger c_{k\downarrow}^{T\dagger}
d_\downarrow d_\downarrow^\dagger c_{k\downarrow}^T) d_\uparrow|\Psi_0\rangle/\Delta_0, \]
where $|\Psi_0\rangle$ denotes the initial and final state wavefunction of Fig.~3 and $\Delta_0$ is the energy unit.
Then, the steady-state nonequilibrium expectation for electron transfer from the tip to sample
should be written as $\langle\Psi_0|{\tilde V}^S\alpha_{k\uparrow}^{\dagger}(operators) d_\uparrow|\Psi_0\rangle/\Delta_0$.
For the reverse movements of Fig.~3(a) and Fig.~3(c), i.e. transfer from the sample to tip,
replacing sample operators by tip operators and vice versa along with changing the hybridization parameter is needed:
\[\langle\Psi_0|{\tilde V}^Tc_{k\uparrow}^{T\dagger}(\alpha_{k\uparrow} d_\uparrow^\dagger c_{k\downarrow}^{T\dagger}
d_\downarrow d_\downarrow^\dagger \alpha_{-k\downarrow}) d_\uparrow|\Psi_0\rangle/\Delta_0 \] and
\[\langle\Psi_0|{\tilde V}^Tc_{k\uparrow}^{T\dagger}(\alpha_{k\uparrow} d_\uparrow^\dagger \alpha_{-k\downarrow}^{\dagger}
d_\downarrow d_\downarrow^\dagger \alpha_{-k\downarrow}) d_\uparrow|\Psi_0\rangle/\Delta_0. \]
Thus, the steady-state nonequilibrium expectation for moving from the sample to tip is written as
$\langle\Psi_0|{\tilde V}^Tc_{k\uparrow}^{T\dagger}(operators) d_\uparrow|\Psi_0\rangle/\Delta_0$.
These steady-state nonequilibrium expectations are bias-independent, which is consistent with
the bias-independence of the local DOS mentioned in section II.
Expectation for six operators appear in the inner products among the basis operators including $\delta$.
One can easily check the orthogonality among those basis operators by observing consecutive appearance of
$d_\uparrow$ ($d_\downarrow$) or $d_\uparrow^{\dagger}$ ($d_\downarrow^{\dagger}$) in the inner products.~\cite{note00}
The six operators used in describing the transports of Fig.~3 appear in the elements of the Liouville matrix given in section IV.

\section{Liouville matrix}
The structure of Liouville matrix ${\bf L}$ sketched in Fig.~2 is composed of two diagonal blocks
sharing the MS and the corner blocks.
Each diagonal block comprises three parts:
an $\infty\times\infty$ diagonal block for the non-interacting reservoir,
a region of hybridization, and a region for the MS.
The corner block contains a tip--sample coherent superposition indicated in orange and a null block
indicated in white due to the lack of direct transition between the tip and sample.
One can reduce the $\infty$-dimensional Liouville matrix to a finite-dimensional matrix
because the tip and sample are represented by diagonal blocks.~\cite{Loewdin,Mujica,ijmp2017}
Thus, the Liouville matrix is ${\bf L}_r={\bf L}_d+{\bf \Sigma}$, where ${\bf L}_d$ is
given by
\begin{eqnarray}
i{\rm\bf L}_d=\left( \begin{array}{c c c c c} 0 & \gamma^T &
-U^T_{j^-} & \gamma^{TS}_S & \gamma^{TS}_A \\ -\gamma^T & 0
& -U^T_{j^+} & \gamma^{TS}_A & \gamma^{TS}_S \\
U_{j^-}^{T*} &  U_{j^+}^{T*} & 0 &  U^{S*}_{j^+} &
U^{S*}_{j^-} \\  -\gamma^{TS}_S & -\gamma^{TS}_A & -U_{j^+}^S  &
0 & -\gamma^S \\
 -\gamma^{TS}_A &  -\gamma^{TS}_S &
 -U_{j^-}^S  & \gamma^S & 0
\end{array} \right), \nonumber
\label{eq:r2r}
\end{eqnarray}
which is the central block indicated by the dashed line in Fig.~2, and
${\bf \Sigma}$ denotes the self-energy matrix created in the process of matrix reduction,
composed of the elements
${\bf \Sigma}_{pq}=\beta_{pq}[\Gamma^T+\Gamma^S(\omega)]/2$
with the coefficients $\beta_{pq}$ coming from the reduction process.~\cite{ijmp2017}
In the atomic limit, where the peaks are given by $\delta$ functions, the Coulomb peaks
position at $\pm U/2$ with $\beta_{pq}=0.25$ in a $T$--$S$ symmetric case except for $\beta_{33}$.
Slight increases occur by correlations.
Hence, we commonly choose
	$\beta_{11} = \beta_{15} = \beta_{55} = 0.252$,
	$\beta_{12} = \beta_{14} = \beta_{25} = \beta_{45} = 0.254$,
	$\beta_{22} = \beta_{24} = \beta_{44} = 0.258$,	$\beta_{33}=1$,
	and $\beta_{13} = \beta_{23} = \beta_{43} = \beta_{53} = 0$
in all strongly correlated systems.~\cite{ijmp2017,Hong-SciRept}

Each matrix element has its own role.
The element $\gamma^{T(S)}$ is given by
$$\gamma^{T(S)}=\frac{\langle\sum_ki(\tilde{V}^Tc_{k\uparrow}^T+\tilde{V}^S\alpha_{k\uparrow})d^\dagger_{\uparrow}
		[j^{-T(S)}_{d\downarrow},j^{+T(S)}_{d\downarrow}]\rangle}
{\langle(\delta j^{-T(S)}_{d\downarrow})^2\rangle^{1/2}\langle(\delta j^{+T(S)}_{d\downarrow})^2\rangle^{1/2}}.$$
The first term of $\gamma^{T}$ describes Fig.~3(c) using
$[j^{-T}_{d\downarrow},j^{+T}_{d\downarrow}]=2i({\tilde V}^T)^2c_{k\downarrow}^{T\dagger}
d_\downarrow d_\downarrow^\dagger c_{k\downarrow}^T$
and the expectation $\langle\Psi_0|{\tilde V}^S\alpha_{k\uparrow}^{\dagger} (operators) d_\uparrow|\Psi_0\rangle/\Delta_0$,
while the second term describes moving in the reverse direction with spin exchange first using
$\langle\Psi_0|{\tilde V}^Tc_{k\uparrow}^{T\dagger}(operators) d_\uparrow|\Psi_0\rangle/\Delta_0$.
On the other hand, Fig.~3(b), which describes Kondo coupling at equilibrium, is the case employing
the first term of $\gamma^{T}$ using the expectation
$\langle\Psi_0|{\tilde V}^Tc_{k\uparrow}^{T\dagger} (operators) d_\uparrow|\Psi_0\rangle/\Delta_0$.
But this is not the dynamics of steady-state nonequilibrium because no electron transport occurs.

The elements $\gamma^{TS}_{S,A}$ are written as
$$\gamma^{TS}_S=\frac{\langle\sum_ki(\tilde{V}^Tc_{k\uparrow}^T+\tilde{V}^S\alpha_{k\uparrow})d^\dagger_{\uparrow}
		[j^{-T}_{d\downarrow},j^{+S}_{d\downarrow}]\rangle}
{\langle(\delta j^{-T}_{d\downarrow})^2\rangle^{1/2}\langle(\delta j^{+S}_{d\downarrow})^2\rangle^{1/2}}$$
and
$$\gamma^{TS}_A=\frac{\langle\sum_ki(\tilde{V}^Tc_{k\uparrow}^T+\tilde{V}^S\alpha_{k\uparrow})d^\dagger_{\uparrow}
		[j^{-T}_{d\downarrow},j^{-S}_{d\downarrow}]\rangle}
{\langle(\delta j^{-T}_{d\downarrow})^2\rangle^{1/2}\langle(\delta j^{-S}_{d\downarrow})^2\rangle^{1/2}},$$
which represent the symmetric and antisymmetric combination of coherent transports between the tip and sample
via MS, respectively: ($T \rightarrow {\mbox MS} \rightarrow S$: first term) +
($T \leftarrow {\mbox MS} \leftarrow S$: second term) for $\gamma^{TS}_S$, and
($T \rightarrow {\mbox MS} \rightarrow S$) $-$ ($T \leftarrow {\mbox MS} \leftarrow S$) for $\gamma^{TS}_A$.
Thus, $\gamma^{TS}_A$ vanishes in equilibrium and $\gamma^{TS}_A=\gamma^{TS}_S$
at steady-state nonequilibrium due to the unidirectional transition of entangled state tunneling.
As a result, the first term of $\gamma^{TS}_{A,S}$ describes Fig.~3(a).
In contrast, the elements $U_{j^\pm}^{T,S}$ are given by
$$U_{j^\mp}^{\alpha}=\frac{iU}{2}\left[\frac{\langle
[n_{d\downarrow},j^{\mp \alpha}_{d\downarrow}]
(1-2n_{d\uparrow})\rangle+\langle\{n_{d\downarrow},\delta j^{\mp
\alpha}_{d\downarrow}\}\rangle}{\langle(\delta j^{\mp
\alpha}_{d\downarrow})^2\rangle^{1/2}}\right],$$
where $\alpha\in T,S$ and $\langle\{n_{d\downarrow},\delta j^{\mp
\alpha}_{d\downarrow}\}\rangle=\langle j^{\mp
\alpha}_{d\downarrow}\rangle(1-2\langle n_{d\downarrow}\rangle)$.
$U_{j^\pm}^{T,S}$ represent the effective on-site Coulomb interaction reflected by fluctuation
in the denominator, and describe the double-occupancy dynamics of
incoherent spins expressed by operators $j^\pm_d$ on the tip or sample side.~\cite{ijmp2017}

\begin{figure}
[b] \vspace*{5cm} \includegraphics{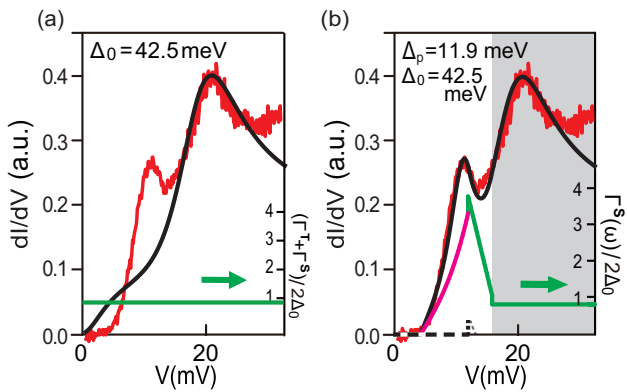} \vspace*{0.0cm}
\caption{ (a) Fitting of the experimental tunneling conductance (red line)
of single-layer FeSe/STO of Ref.~2 using flat DOS (green line).
The black line is obtained using the values in Table I and constant DOS of
$\Gamma^T=0.24\Delta_0$ and $\Gamma^S=1.66\Delta_0$ with energy unit $\Delta_0$.
(b) Fitting of the same experimental tunneling conductance in (a) by varying sample DOS $\Gamma^S(\omega)$ in the low-energy region of
$\omega\lesssim 14$ meV.
The sample DOS (magenta+green line) is now composed of a linear-like part (magenta line) and
suppressed $s$-wave DOS (black dashed line). The grey area indicates the region of flat DOS.
 }
\end{figure}

\section{Results}
With the reduced Liouville matrix ${\bf M}_r$ defined by ${\bf M}_r=z{\bf I}-i{\bf L}_r$ with $z=-i\omega+0^+$,
we finally arrive at the expression for tunneling conductance,
\begin{equation}
    \frac{dI}{dV}=\left. \frac{2e^2}{h}\left[\frac{\Gamma^T\Gamma^S(\omega)}
    {\Gamma^T+\Gamma^S(\omega)}\right]
    {\rm Re}\left({\bf M}_r^{-1}\right)_{dd}\right|_{\hbar\omega=eV}.
	\label{Tcond}
\end{equation}
Equation (\ref{Tcond}) gives the tunneling conductance as a function of
$\Gamma^S(\omega)$ that is proportional to $D(\omega)$.
Note that $\Gamma^S(\omega)$ appears in both the self-energy matrix ${\bf \Sigma}$
and the front factor.
As it is practically difficult to calculate $\gamma$s and $U_{j^\pm}^{T,S}$ for the
Hamiltonian given in Eq. (\ref{Hamil2}), we leave them as free parameters.
Despite this, though, we are able to infer the relative magnitudes of $\gamma$s by considering the degrees of fluctuation
in the metallic tip and superconductor sample.
It is reasonable to surmise that fluctuation
$\langle(\delta j^{\pm T}_{d\downarrow})^2\rangle^{1/2}$ in the tip is bigger than
$\langle(\delta j^{\pm S}_{d\downarrow})^2\rangle^{1/2}$ in the superconductor,
which results in $\gamma^T<<\gamma^S$ because fluctuation appears in the denominator.
We apply the same reasoning in choosing the values of $\mathrm{Re}[U^T_{j^+}]$ and $\mathrm{Re}[U^S_{j^+}]$.
However, the steady current condition gives $\mathrm{Re}[U^T_{j^-}]=\mathrm{Re}[U^S_{j^-}]$.
We set the matrix element values given in Table I according to this analysis and the condition for
entangled state tunneling $\gamma_A^{TS}=\gamma_S^{TS}$.

Now, we demonstrate a typical procedure to obtain the DOS function using the STS data for
single-layer FeSe/STO reported in Ref.~2.
First, we fit the high-bias peak (major peak) of the tunneling conductance using the matrix
elements given in Table~I and constant sample DOS for both tip and sample, which gives
Fig.~4(a).
All free parameters except $\Gamma^S(\omega)$ are then fixed with this fitting procedure.
Figure~4(a) provides us with the surprising conclusion that the major peak located
at 21 meV is not related to the sample DOS. It is just the effect of nonequilibrium entangled state tunneling.
Next, we try to fit the low-bias part of the tunneling conductance by varying
$\Gamma^S(\omega)$ in Eq. (\ref{Tcond}) to ultimately obtain $D(\omega)$ (magenta+green line)
and a well-fitting tunneling conductance (black line) simultaneously, as shown in Fig.~4(b).
Quantitative agreement up to the high-bias peak is almost exact.
The peak of the sample DOS in Fig.~4(b) is located at 11.9 meV, which coincides
with the maximum of the ARPES energy distribution curve reported in Refs.~11 and 12.

Before presenting the result of our study on extracting the $s$-wave DOS functions in
Ba$_{1-x}$K$_x$Fe$_2$As$_2$, we analyze the DOS function given in Fig.~4(b).
The line shape of the DOS function will be mostly formed by the band at the M point because
the band at the $\Gamma$ point is suppressed below the Fermi level, as discussed earlier.
The magenta line in Fig.~4(b) represents the contribution by the band at the M point, and
the short upright at the top seems to be a weak indirect contribution from the band at the $\Gamma$ point;
we separate it using the black dashed line near the horizontal axis.
We conclude that the sample DOS is composed of a linear-type DOS by the band at the M point,
and the very much suppressed $s$-wave DOS by the band at the $\Gamma$ point.

\twocolumngrid
\begin{center}
\begin{table}[t]
\caption{\textbf{Values of matrix elements.}
	The values of the matrix elements appearing in ${\bf L}_d$
	used to obtain Figs.~4 and 5. We set $\mathrm{Im}[U_{j^\pm}]=0$, and all values are
	in units of $\Delta_0$.}
    \setlength{\tabcolsep}{5pt}
    \begin{tabular}{c|cccccc}
    \hline
 Fig. & $\gamma^T$ & $\gamma^S$ & $\gamma^{TS}_{A,S}$ &
        $\mathrm{Re}[U^T_{j^+}]$ & $\mathrm{Re}[U^S_{j^+}]$ &
        $\mathrm{Re}[U^{T,S}_{j^-}]$ \\
\hline
4 & 0.01 & 0.5 & 0.5 & 0.9  & 6.3 & 4.5   \\
5 & 0.01 & 0.45 & 0.45 & 0.99 & 3.97 & 2.37 \\
\end{tabular}
\end{table}
\end{center}
\twocolumngrid

Finally, we present the result for Ba$_{1-x}$K$_x$Fe$_2$As$_2$ in Fig.~5.
The value of the experimental tunneling conductance at zero bias is exceptionally non-vanishing.~\cite{Wray}
We judge that this non-vanishing at zero bias is not an intrinsic phenomenon, and therefore
we study the STS data by considering the minimum at zero bias as the origin of tunneling conductance.

In Fig.~5, we show the extracted sample DOS (magenta+orange+green line) which contains
two $s$-wave DOS (black dashed lines indicated by $\Gamma_1$ and $\Gamma_2$),
although one of them ($\Gamma_1$) is not discernible.
The peaks of the $s$-wave DOS are located at 5.26 meV and 14.5 meV, which are well matched
with ARPES results.~\cite{Wray,Shimojima}
These $s$-wave DOS correspond to two different Fermi surfaces around the $\Gamma$ point:
the outer Fermi surface ($\Gamma_2$) produces a pronounced $s$-wave peak at lower energy, while
the inner Fermi surface ($\Gamma_1$) generates a lower $s$-wave peak at higher energy.
In Ref.~24, an additional Fermi surface exists between $\Gamma_1$ and $\Gamma_2$, but the effect of
this middle Fermi surface is not prominent in the tunneling data of Ref.~7.
In contrast, a linear DOS (magenta line) is formed within the major $s$-wave gap.
This linear DOS is obviously produced by the band at the M point, according to the analysis
in Fig.~4 where the band at the $\Gamma$ point is suppressed below the Fermi level.
The existence of extra DOS (magenta line) within the $s$-wave gap causes a decrease in superconducting
transition temperature $T_c$, as shown by $\Delta_{p2}=5.26$ meV and $T_c=3.2$ meV in Fig.~5.

\begin{figure}
[t] \vspace*{8cm} \includegraphics{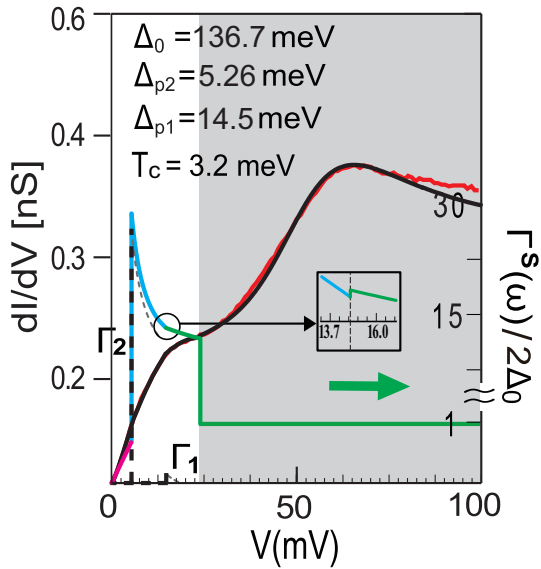} \vspace*{0.0cm}
\caption{Fitting of the tunneling conductance of Ba$_{1-x}$K$_x$Fe$_2$As$_2$
reported in Ref.~7 with the analyzed sample DOS.
The solid black line is our theoretical curve, and the red line is the experimental data.
The sample DOS is composed of a linear part (magenta line) at low bias, blue, and green lines.
Two $s$-wave DOS (black dashed lines: $\Gamma_1$ and $\Gamma_2$) are shown with gray dotted tails
for ease of viewing.
The inset magnifies the crossing area.
The grey area indicates the region of flat DOS.
  }
\end{figure}

\section{Discussion}
We have revealed ordinary $s$-wave DOS of Fe-based superconductor Ba$_{1-x}$K$_x$Fe$_2$As$_2$
hidden in the tunneling conductance.
We adopted the Liouville approach in which a complete set of basis vectors (operators) can be determined.
The infinite-dimensional Liouville matrix is reduced to a $5\times 5$ matrix yielding
five general peaks in local DOS $\rho_d(\omega)$:
one coherent peak at the Fermi level, two additional low-energy coherent peaks,
and two incoherent Coulomb peaks.
The three coherent peaks can be seen in various mesoscopic quantum impurity systems.~\cite{Cronenwett,Sarkozy-PRB79}
However, the zero-bias peak is suppressed in the case of strong asymmetry in coupling
strength, i.e., $\gamma^T\ll\gamma^S$, or when sample DOS has a gap or vanishes at the Fermi level.
In contrast, the two incoherent Coulomb peaks position far outside the concerned energy range.
Thus, the two additional low-energy coherent peaks remain alone in the tunneling conductance of
correlated superconductors.
The low-energy coherent peak appears as a major peak in STS data and is misunderstood
as a superconducting peak.~\cite{Hong-SciRept}
We stress here that the major peak shown in the STS data is solely an effect of nonequilibrium,
being a unique feature of entangled state tunneling under bias.
The sample DOS of a correlated superconductor is not usually prominent in the STS line shape;
therefore, a theoretical formula relating tunneling conductance and sample DOS is necessary to
extract the sample DOS from tunneling data.
Equation (\ref{Tcond}) plays this role.
To extract a reliable sample DOS using Eq. (\ref{Tcond}), high-quality STS data are required.
One may see the effect of the middle Fermi surface reported in Ref.~24
if a better STS line shape is provided.

\centerline{\bf ACKNOWLEDGMENTS}

This work was supported by Project Code (2017R1D1A1A02017587) and partially supported by a KIAS grant funded by MSIP.



\begin{thebibliography}{0}
\bibitem{note00} Orthogonality condition was used without proof in Ref.~14.
\end{thebibliography}
\end{document}